\documentclass[twocolumn,pra,aps,showpacs,superscriptaddress,floatfix,showkeys,nofootinbib,10pt]{revtex4-1}

\usepackage{amsmath,amsfonts,amssymb,amsthm}
\usepackage{graphicx,mathtools,color,enumerate}
\usepackage{bm}
\usepackage[makeroom]{cancel}

\newcommand{\ket}[1]{ | \, #1 \rangle} \newcommand{\bra}[1]{ \langle #1 \, |} 
\newcommand{\proj}[1]{\ket{#1}\bra{#1}} 
\newcommand{\kb}[2]{\ket{#1}\bra{#2}}

\newcommand{\E}[1]{ \langle #1 \, \rangle} 
\newcommand{\Ab}[1]{ \left| #1 \, \right|} 
\newcommand{\be}{\begin{equation}} \newcommand{\ee}{\end{equation}}
\newcommand{\ba}{\begin{aligned}} \newcommand{\ea}{\end{aligned}}

\DeclareMathOperator{\Tr}{Tr}
\DeclareMathOperator{\Det}{Det}

\newcommand{\me}{\mathrm{e}}

\newcommand\bigforall{\mbox{\Large $\mathsurround=0pt\forall$}} 
\newcommand\bigexists{\mbox{\Large $\mathsurround=0pt\exists$}}

\DeclareMathOperator{\probI}{p_i}

\newtheorem{prop}{Proposition}
\newtheorem{cor}{Corollary}
\newtheorem{lemma}{Lemma}
\newtheorem{obs}{Observation}

\begin{document}

\title{System information propagation for spin structures}

\author{P.~Mironowicz} \email{piotr.mironowicz@gmail.com} \affiliation{Department of Algorithms and System Modeling, Faculty of Electronics, Telecommunications and Informatics, Gda\'nsk University of Technology} \affiliation{National Quantum Information Centre in Gda\'nsk, 81-824 Sopot, Poland}
\author{P.~Nale\.zyty} \affiliation{Independent Researcher}
\author{P.~Horodecki} \email{pawel@mif.pg.gda.pl} \affiliation{Faculty of Applied Physics and Mathematics, Gda\'nsk University of Technology, 80-233 Gda\'nsk, Poland} \affiliation{National Quantum Information Centre in Gda\'nsk, 81-824 Sopot, Poland}
\author{J.~K.~Korbicz} \affiliation{Faculty of Applied Physics and Mathematics, Gda\'nsk University of Technology, 80-233 Gda\'nsk, Poland} \affiliation{National Quantum Information Centre in Gda\'nsk, 81-824 Sopot,
Poland}

\date{\today}

\begin{abstract}
We study in details decoherence process of a spin register, coupled to a spin environment. We use recently developed methods of information transfer study in open quantum systems to analyze information flow between the register and its environment. We show that there are regimes when not only the register decoheres effectively to a classical bit string, but this bit string is redundantly encoded in the environment, making it available to multiple observations. This process is more subtle than in a case of a single qubit due to possible presence of protected subspaces: Decoherence free subspaces and, so called, orthogonalization free subspaces. 
We show that this leads to a rich structure of coherence loss/protection in the asymptotic state of the register and a part of its environment. We formulate a series of examples illustrating these structures. 
\end{abstract}

\keywords{decoherence, quantum Darwinism, state broadcasting}

\maketitle

\section{Introduction}

In our previous paper~\cite{ourPrevious} we investigated the process of formation of the, so-called, Spectrum Broadcasting Structure (SBS), see~\eqref{eq:broadcastState} below for the definition, in the spin-spin model \cite{Zurek82,Zurek05,SchlossauerBook}. 
Our main result was the description of how the interaction of such a form leads to the objectivization of the information about the central system. The aim of this paper is to develop the ideas sketched previously. We also consider more explicit examples of the objectivization process.

The problem how knowledge, being a \textit{classical} (not quantum) resource, is protruding out of quantum regime, has been deeply investigated by W.~H.~\.Zurek \textit{et al.} in a series of works~\cite{Zurek81,Zurek82,Zurek93,Zurek99,Zurek01,Zurek03,Zurek04,Zurek05} (see~\cite{ZurekRMP,SchlosshauerRMP} for an overview) leading to the concept of the \textit{quantum Darwinism}~\cite{Zurek06,ZurekNature}.

The essence of the quantum Darwinism is the statement that this information about a measured system which is being efficiently proliferated into different parts of the environment, and in consequence \textit{redundantly imprinted and stored} in them, becomes objective. Each part contains an almost complete classical information about the system. This redundancy is crucial for the objectivity~\cite{Zurek06,Zurek09,Zurek10,ZurekNature,Zurek14,myObj,Brandao15}, as parts may be accessed independently by many observers gaining the same or similar knowledge. See~\cite{Brunner08,Bruke10} for experimental demonstration of effects of the quantum Darwinism.

The problem how some information is being distributed in many copies by an intrinsically quantum mechanism is highly nontrivial, due to the no-cloning theorem~\cite{nocloning1,nocloning2} forbidding the direct copying of a state. What is more, even a weaker form of copying of quantum states, the so-called state broadcasting, is not always possible~\cite{broadcast96,nolocalbroadcast}.

We note here that even though the term \textit{objectivity} is widely used in this context, in order to be more precise one should use the word \textit{intersubjectivity}, in the meaning of Ajdukiewicz~\cite{Ajdukiewicz49,AjdukiewiczEng}, instead. It has been shown recently~\cite{myObj} that the emergence of the classical and objective properties in this spirit is due to a form of information broadcasting (similar to the so-called spectrum broadcasting~\cite{broadcast12}) leading to the creation of a specific quantum state structure between the system and a part of its environment, namely the SBS mentioned above. The SBS ensures different observers a perfect access to some property of the observed system.

Throughout this paper we consider how the intersubjectivity of the state of the central system can emerge in the spin-spin model~\cite{Zurek82,Zurek05,SchlossauerBook,Zurek2016} in the quantum measurement limit.

The work is organized as follows. First, in sec.~\ref{sec:broadCentral} we introduce basic terms and recapitulate the results of~\cite{ourPrevious} in a more detailed and instructive way. The results concerns the case with a single central spin surrounded by a number of environmental spins. In sec.~\ref{sec:spinSpin} we define and calculate the so-called orthogonalization and decoherence factors for this model.

Next, in sec.~\ref{sec:asympt}, we investigate the asymptotic behaviour of the orthogonalization and decoherence processes in spin models using the Weak Law of Large Numbers (LLN). We develop some tools which are useful for the analysis of the quasi--periodic functions often occurring in similar models~\cite{ourPrevious,Zurek81,Zurek05,Nalezyty,Tuziemski17}. In sec.~\ref{sec:register} we consider a more involved example of a central system within a spin environment, namely a register of spins. Finally, in sec.~\ref{sec:DecOrthFreeExamples}, we give examples of orthogonalization--free and decoherence--free setups.

\section{Spectrum Broadcast Structures in the single central spin model}
\label{sec:broadCentral}

Let us consider a central system~$S$ interacting with $M$-partite environment, with $fM$ being the number of parts under observation, $f \in (0,1)$. We say that the joint state of $S$ and $fM$ parts of the environment, $\varrho_{SBS}(t)$, constitutes an SBS if it is of the form:
\be 
	\label{eq:broadcastState}
	\varrho_{SBS}(t) \equiv \sum_{i} \probI(t) \proj{i} \otimes \bigotimes_{k=1}^{fM} \varrho^{(k)}_{i}(t), 
\ee 
with $\varrho^{(k)}_{i}(t)$ and $\varrho^{(k)}_{j}(t)$ for $i \neq j$ having orthogonal supports (meaning perfect distinguishability with a single measurement). By measuring the state $\varrho^{(k)}_{i}(t)$ any of the local observers extracts the same information about the state of the system, i.e. the index $i$, without disturbing it (after forgetting the results).

A convenient measure of orthogonality of a pair of states, $\varrho_{+}$ and $\varrho_{-}$, is the so-called generalized fidelity, or overlap~\cite{Fuchs99},
\be
	B(\varrho_{+}, \varrho_{-}) \equiv \Tr\sqrt{ \sqrt{\varrho_{+}} \varrho_{-} \sqrt{\varrho_{+}} }.
\ee
This function is multiplicative, namely it has the following useful property:
\be
	\label{eq:Bprod}
	B \left( \varrho^{(1)}_{+} \otimes \varrho^{(2)}_{+}, \varrho^{(1)}_{-} \otimes \varrho^{(2)}_{-} \right) = B \left( \varrho^{(1)}_{+}, \varrho^{(1)}_{-} \right) \cdot B \left( \varrho^{(2)}_{+}, \varrho^{(2)}_{-} \right).
\ee

\subsection{Dynamics}

As mentioned above, in our considerations we assume the quantum measurement limit, meaning that the central interaction Hamiltonian dominates the dynamics. The assumption that the environmental subsystem do not interact simplifies the analysis significantly and is a common practice. Thus the evolution is governed by a Hamiltonian of the general von Neumann 
measurement form:
\be
	\label{eq:generalHint}
	H_{int} = X \otimes \sum_{k=1}^M Y_k.
\ee
The resulting evolution is given by the following unitary operator:
\be
	\label{eq:Uqml}
	U(t) \equiv \exp \left( - i H_{int} t \right) = \sum_i \proj{i} \otimes \bigotimes_{k=1}^{M} U^{(k)}_{i}(t),
\ee
where $U^{(k)}_{j}(t) \equiv \exp \left( - i x_j Y_k t \right)$, and $\sum_{i=1}^{d_S} x_i \ket{i}\bra{i} \equiv X$. Let $\varrho(0)$ denote the initial state of the central system together with the environment, and $\varrho(t)$ the evolved state:
\be
	\varrho(t) = U(t) \varrho(0) U^{\dagger}(t).
\ee

Within the context of the quantum measurement limit we assume that the initial state is in a product form:
\be
	\label{eq:initialStateProduct}
	\ba
	\varrho(0) =& \varrho_{S}(0) \otimes \bigotimes_{k=1}^{M} \varrho^{(k)}(0) \\
	=& \varrho_{S}(0) \otimes \bigotimes_{k=1}^{fM} \varrho^{(k)}(0) \otimes \bigotimes_{k=1}^{(1-f)M} \varrho^{(fM+k)}(0).
	\ea
\ee
Let us note that if we discard the $(1-f)M$ parts of the environment we get a partially traced state:
\be
	\label{reduced} 
	\ba 
		\varrho^{(fM)}(t) \equiv & {\Tr}_{(1-f)M} \varrho(t) = \sum_{i=1}^{d_S} \sigma_{i} \proj{i} \otimes \bigotimes_{k=1}^{fM} \varrho^{(k)}_{i}(t) + \\ 
		& \sum_{i \neq i'} \left[ \sigma_{i i'} \prod_{k=1}^{(1-f)M} \gamma_{i i'}^{(k)}(t) \right] \kb{i}{i'} \otimes \bigotimes_{k=1}^{fM} \varrho_{i, i'}^{(k)}(t),
	\ea
\ee 
with
\begin{subequations}
	\label{eq:varrhoI}
	\be
		\varrho_{i, i'}^{(k)}(t) \equiv U_{i}^{(k)}(t) \varrho^{(k)}(0) U_{i'}^{(k) \dagger}(t),
	\ee
	\be
		\varrho_{i}^{(k)}(t) \equiv \varrho_{i,i}^{(k)}(t),
	\ee
\end{subequations}
and $\sigma_{i i'} \equiv \bra{i} \varrho(0) \ket{i'}$, $\sigma_i \equiv \sigma_{i,i}$, and $\gamma_{ij}^{(k)}(t) \equiv \Tr\left[ \varrho_{i,j}^{(k)}(t) \right]$. The products:
\be
	\label{eq:gammaij}
	\gamma_{i i'}(t) \equiv \prod_{k=1}^{(1-f)M} \gamma_{i i'}^{(k)}(t),
\ee
are called the decoherence factors.

A simple example of a Hamiltonian of the form~\eqref{eq:generalHint} is the spin-spin model, one of the canonical models of decoherence~\cite{Zurek81,Zurek82,Zurek05,Zurek05b,SchlossauerBook}. The interaction Hamiltonian for a central spin surrounded by $N$ spins reads:
\be
	\label{eq:simpleSpinSpin}
	H_{int} = \frac{1}{2} \sigma_z \otimes \sum_{j=1}^N g_j \sigma_z^{(j)},
\ee
where $g_j$ are coupling constants and Pauli matrices $\sigma_z^{(j)}$ are acting on the space of the $j$-th spin. However, it has been only briefly analyzed from the point of view of information transfer in terms of SBS~\cite{ourPrevious} and here we provide such an analysis, generalizing the original model to that of a $K$-qubit register: 
\be
	\label{eq:registerHint}
	H_{int} = \frac{1}{2} \sum_{i=1}^{K} \tilde{\sigma}_z^{(i)} \bigotimes \sum_{j=1}^{N} g_{ij} \sigma_z^{(j)},
\ee
where we use $\tilde{\sigma}_z^{(i)}$ to denote the $\sigma_z$ matrix acting on the $i$-th register, and $\sigma_z^{(j)}$ acts on $j$-th of the environmental spins.

\subsection{Decoherence and overlap factors}
\label{sec:spinSpin}

We first recall our analysis~\cite{ourPrevious} of the single central spin case~\eqref{eq:simpleSpinSpin}, which will be the basis for studying the more general spin register model~\eqref{eq:registerHint}.

We divide the $N$ environmental spins into $M$ arbitrary disjoint parts, $mac_1, \cdots, mac_M$. Without loss of generality, we may assume that the spins are labeled in such a way, that each part contain consecutive spins. The $k$-th part is identified with a subset of indices $mac_k \subseteq \{1, \cdots, N\}$, and is called a macrofraction if $\Ab{mac_k}$ scales with $N$~\cite{myPRL}, introducing thus an environmental coarse-graining.

As noted previously, the first $fM$ macrofractions are being observed. Let $Obs$ denote the set of all observed spins, i.e. $Obs \equiv \bigcup_{k=1}^{fM} mac_k$. We discard the remaining spins as unobserved. The interaction~\eqref{eq:simpleSpinSpin} can be rewritten in the following way:
\be
	\label{eq:HintMacrof}
	H_{int} = \frac{1}{2} \sigma_z \otimes \left[ \sum_{k=1}^{fM} \left( \sum_{j \in mac_k} g_j \tilde{\sigma}_z^{(j)}\right) + \sum_{j \notin Obs} g_j \sigma_z^{(j)} \right].
\ee

We use the following parametrization of the initial state of spins,
\be
	\varrho(0) \equiv
	\begin{bmatrix}
		\sigma_{+} & \sigma_{+-} \\
		\sigma_{+-}^{\dagger} & \sigma_{-}
	\end{bmatrix}
	\otimes
	\bigotimes_{j=1}^{N} \rho^{(j)}(0),
\ee
cf.~\eqref{eq:initialStateProduct} with the $SU(2)$ Euler angles \cite{EulerSUN}:
\be
	\rho^{(j)}(0) = R^{(j)} D^{(j)} R^{(j)\dagger},
\ee
where
\begin{subequations}
	\be
		\label{eq:Dlambda}
		D^{(j)} \equiv \text{diag}(\lambda_j, 1 - \lambda_j),
	\ee
	\be
		\label{eq:Euler}
		\ba
			R^{(j)} & \equiv e^{i \frac{\alpha}{2} \sigma_z} e^{i \frac{\beta}{2} \sigma_y} e^{i \frac{\gamma}{2} \sigma_z} \\
			&= \begin{bmatrix}
				e^{\tfrac{i}{2}(\alpha_j + \gamma_j)}\cos\frac{\beta_j}{2} & e^{\tfrac{i}{2}(\alpha_j - \gamma_j)}\sin\frac{\beta_j}{2} \\
				-e^{-\tfrac{i}{2}(\alpha_j - \gamma_j)}\sin\frac{\beta_j}{2} & e^{-\tfrac{i}{2}(\alpha_j + \gamma_j)}\cos\frac{\beta_j}{2}
			\end{bmatrix}.
		\ea
	\ee
\end{subequations}
For the interaction~\eqref{eq:HintMacrof} we have
\be
	U_{\pm}^{(j)} = U_{\mp}^{(j)\dagger} = \exp \left( \mp \frac{i}{2} g_j t \sigma_z \right) = 
		\begin{bmatrix}
			e^{\mp \frac{i}{2} g_j t} & 0 \\
			0 & e^{\pm \frac{i}{2} g_j t}
		\end{bmatrix}.
\ee
With this parametrization we see from~\eqref{eq:varrhoI} that the explicit formula for the states $\rho_{\pm}^{(j)}(t)$ and $\rho_{\pm,\mp}^{(j)}(t)$ is the following:
\begin{subequations}
	\be
		\ba
			\rho_{\pm}^{(j)}(t) &= U_{\pm}^{(j)}(t) R^{(j)} D^{(j)} R^{(j)\dagger} U_{\pm}^{(j)\dagger}(t)\\
			& = \frac{1}{2}
			\begin{bmatrix}
				1+\zeta_j & e^{\mp i g_j t} \vartheta_j \\
			 e^{\pm i g_j t} \vartheta_j^{*} & 1-\zeta_j
			\end{bmatrix},
		\ea
	\ee
	\be
		\ba
			\rho_{\pm,\mp}^{(j)}(t) &= U_{\pm}^{(j)}(t) R^{(j)} D^{(j)} R^{(j)\dagger} U_{\mp}^{(j)\dagger}(t)\\
			& = \frac{1}{2}
			\begin{bmatrix}
				(1+\zeta_j) e^{\pm i g_j t} & \vartheta_j \\
				\vartheta_j^{*} & (1-\zeta_j) e^{\mp i g_j t}
			\end{bmatrix}.
		\ea
	\ee
\end{subequations}
Here 
\begin{eqnarray}
&& \vartheta_j \equiv - \tfrac{1}{2} (2 \lambda_j - 1) \sin{\beta_j},\\
&& \zeta_j \equiv (2 \lambda_j - 1) \cos{\beta_j}.\label{zeta}
\end{eqnarray}

After those preparations we are ready to calculate the two central functions of our analysis: The appropriate decoherence and overlap factors.
Their importance stems from the following crucial result \cite{ourPrevious}: The optimal trace-norm distance 
$\epsilon_{SBS}$ of the actual partially traced state~\eqref{reduced} to an ideal SBS is bounded by:
\be\label{error}
	\epsilon_{SBS} \leq \Ab{\sigma_{+-}} \cdot \Ab{\gamma(t)} + \sqrt{\sigma_{+} \sigma_{-}} \sum_{k=1}^{fM} B^{(k)}(t),
\ee
where $B^{(k)}(t) \equiv B\left[\varrho_{+}^{(k)}(t), \varrho_{-}^{(k)}(t)\right]$.

The decoherence factor $\gamma(t) \equiv \gamma_{-+}(t)$, cf.~\eqref{eq:gammaij}, is well known in this model and has been calculated in ~\cite{Zurek05}. It reads:
\be
	\label{eq:gamma2}
	\ba
		\gamma(t) &= \prod_{j \notin Obs} \left[ \Tr \left( \exp \left( i g_j t \sigma_z \right) \rho^{(j)}(0) \right) \right] \\
		&= \prod_{j \notin Obs} \left[ \frac{1+\zeta_j}{2} \me^{i g_j t} + \frac{1-\zeta_j}{2} \me^{-i g_j t} \right] \\
		&= \prod_{j \notin Obs} \left[ \cos \left(g_j t\right) + i \zeta_j \sin \left(g_j t\right) \right].
	\ea
\ee

We now calculate the overlap $B \left( \rho_{+}^{(j)}(t), \rho_{-}^{(j)}(t) \right)$. Let us define:
\be
	\label{eq:M}
	\mathbf{M}^{(j)} \equiv \sqrt{D^{(j)}} R^{(j)\dagger} U_{-}^{(j)2} R^{(j)} D ^{(j)} R^{(j)\dagger} U_{+}^{(j)2} R^{(j)} \sqrt{D^{(j)}}.
\ee
After pulling some of the unitary operators out of the square roots and using the cyclic property of the trace we obtain:
\be
	B \left( \rho_{+}^{(j)}(t), \rho_{-}^{(j)}(t) \right) = \Tr\sqrt{\mathbf{M}^{(j)}}.
\ee
For a $2 \times 2$ matrix $\mathbf{M}^{(j)}$ the eigenvalues $M^{(j)}_{\pm}$ satisfy:
\be
	\label{eq:Mpm}
	M^{(j)}_{\pm} = \frac{1}{2} \left[ \Tr\mathbf{M}^{(j)} \pm \sqrt{(\Tr\mathbf{M}^{(j)})^2 - 4\Det\mathbf{M}^{(j)}} \right].
\ee
Straightforward calculations show that for $\mathbf{M}^{(j)}$ given by~\eqref{eq:M} we have:
\be
	\Tr \mathbf{M}^{(j)} = \lambda_j^2 + (1-\lambda_j)^2 - \left( 2\lambda_j - 1 \right)^2 \sin^2 \beta_j \sin^2(g_j t),
\ee
and $\Det\mathbf{M}^{(j)} = \lambda_j^2 (1-\lambda_j)^2$. From~\eqref{eq:Mpm} it follows that $M^{(j)}_{+}(t) M^{(j)}_{-}(t) = \Det\mathbf{M}^{(j)}$, and thus:
\be
	\label{eq:Bj}
	\ba
		B & \left( \rho_{+}^{(j)}(t), \rho_{-}^{(j)}(t) \right) = \Ab{ \sqrt{M^{(j)}_{+}(t)}+\sqrt{M^{(j)}_{-}(t)} } \\
		&= \left( M^{(j)}_{+}(t) + M^{(j)}_{-}(t) + 2 \sqrt{M^{(j)}_{+}(t) M^{(j)}_{-}(t)} \right)^{\frac{1}{2}} \\
		&= \sqrt{\Tr\mathbf{M}^{(j)}(t) + 2\sqrt{\Det\mathbf{M}^{(j)}(t)}} \\
		&= \sqrt{1 - (2 \lambda_j - 1)^2 \sin^2 \beta_j \sin^2(g_j t)}.
	\ea
\ee

For a particular macrofraction, $mac_k$, we define the macrofraction states: 
\be
	\label{eq:rhoMac}
	\varrho_{\pm}^{(k)}(t) \equiv \bigotimes_{j \in mac_k} \rho_{\pm}^{(j)}(t).
\ee
We now calculate the overlap:
\be
	B^{(k)}(t) = B\left[\varrho_{+}^{(k)}(t), \varrho_{-}^{(k)}(t)\right].
\ee
Using iteratively the property~\eqref{eq:Bprod}, from~\eqref{eq:Bj} we obtain~\cite{ourPrevious}:
\be
	\label{eq:B}
	\ba
		B^{(k)}(t) &= \prod_{j \in mac_k} \Ab{ \sqrt{M^{(j)}_{+}(t)}+\sqrt{M^{(j)}_{-}(t)} } \\
		& = \prod_{j \in mac_k} \sqrt{1 - (2 \lambda_j - 1)^2 \sin^2 \beta_j \sin^2(g_j t)}.
	\ea
\ee

For an individual spin the functions within products in~\eqref{eq:B} and~\eqref{eq:gamma2} are periodic in time with the frequency $g_j$. The coarse-graining of the environment into macrofractions together with random couplings turns the above functions into quasi-periodic ones. We give a more involved analysis of such functions in Appendix~\ref{sec:avgPeriodic}.

\subsection{Asymptotic behavior of $B(t)$ and $\gamma(t)$}
\label{sec:asympt}

Further analysis employs approximation methods. We now assume that all the macrofractions (the observed and the unobserved) are large enough so that we can use the Law of Large Numbers (LLN)~\cite{LLN}, stating that a sample average converges to the expected value. This approach is to be compared with~\cite{Zurek05}, where the Central Limit Theorem was used in the analysis of the decoherence factor~\eqref{eq:gamma2}, owing to the fact that the later can be represented as a Fourier transform of a probability measure on sums of independent random variables.

In the light of this, the use of LLN is perhaps a bit less natural when studying the decoherence factor, as we must introduce 'by hand' some distribution of both the coupling constants $g_j$ and the initial state parameters. Nonetheless it allows to study the overlap~\eqref{eq:B} and the decoherence factor~\eqref{eq:gamma2}, at least approximately.
We choose the following probability distributions:
\begin{enumerate}[i)]
	\item the initial state Euler angles $(\alpha_j,\beta_j,\gamma_j)$~\eqref{eq:Euler} are distributed with the $SU(2)$ Haar measure;
	\item the initial state eigenvalue $\lambda_j$~\eqref{eq:Dlambda} is distributed according to the eigenvalue part of the Hilbert-Schmidt measure~\cite{HSmeas}:
	\be
		P_{HS}(\lambda_j)\equiv 3 (2 \lambda_j - 1)^2;
	\ee
	\item the coupling constants $g_j$ are distributed with any continuous measure of a non-zero and finite second moment, $\overline{g^2} \equiv \E{g^2_j} > 0$.
\end{enumerate}
One can then easily calculate the following averages over these distributions: $\E{ \sin^2 \beta_j} = \frac{2}{3}$, $\E{\cos^2 \beta_j} = \frac{1}{3}$, $\E{(2 \lambda_j - 1)^2} = \frac{3}{5}$.

Let us consider a particular macrofraction, $mac = mac_k$ for some $k$, and denote $B(t) \equiv B^{(k)}(t)$, $N_{mac} \equiv \Ab{mac}$, $N_{dis} \equiv N - \Ab{Obs}$. Under the assumptions i)--iii) we show in Appendix~\ref{largev} that if the size of macrofractions and the environment is large enough, then with high probability the values of $B(t)$ and $\gamma(t)$ are small, as stated in the following:

\begin{prop}
\label{prop:LLN}
For any $\epsilon, t > 0$ and for $N_{mac}$ and $N_{dis}$ large enough with probability at least $1 - \epsilon$ we have:
\begin{subequations}
	\label{eq:propLLN}
	\be
		B(t) \leq \epsilon,
	\ee
	\be
		\Ab{\gamma(t)}^2 \leq \epsilon.
	\ee
\end{subequations}
\end{prop}
Thus from~\eqref{error} the partially traced state~\eqref{reduced} approaches in the trace norm the SBS form.

More can be said if one assumes a concrete $g_j$ distribution. As an illustration of our methods, let us assume $g_j$ to be i.i.d.
with the uniform distribution over $[0,1]$ \cite{Zurek05}. We then have the following bounds in the short-time regime \cite{ourPrevious}:
 \begin{subequations}
	\label{shorttime}
	\be
		B(t) \leq \exp \left[-\frac{1}{5}N_{mac} \overline{g^2}t^2 \right],
	\ee
	\be
		\Ab{\gamma(t)}^2 \leq \exp \left[-\frac{4}{5} N_{dis} \overline{g^2}t^2\right].
	\ee
\end{subequations}
This initial of course decay by no means guarantees that the functions cannot revive. In finite dimensional setting they will in fact revive, but increasing the environment size one can make the revivals highly unlikely as per Proposition~\ref{prop:LLN}. Indeed, one can show that asymptotically the following bounds hold \cite{ourPrevious}:
\begin{subequations}
	\label{eqs:BgammaExpBoundsUniformLargeT}
	\be
		\lim_{t \to \infty} B(t) \leq \exp \left[-\frac{1}{10} N_{mac} \right],
	\ee
	\be
		\lim_{t \to \infty} \Ab{\gamma(t)}^2 \leq \exp \left[-\frac{2}{5} N_{dis} \right].
	\ee
\end{subequations}
All the above results are derived in Appendix~\ref{uniform} for completeness.

Above we considered formulas for orthogonalization and decoherence factors for sufficiently large environments at a given moment of time. In contrast, we can also ask about long time averages of $B(t)$ and $\gamma(t)$ for a given finite size of the environment, as investigated in the Proposition~\ref{prop:timeAvg} below.

We say that a set of numbers $\{\alpha_i\}_{i=1}^{N} \subset \mathbb{R}$ is impartitionable if for any $\varsigma \in \{\pm 1\}^{N}$ we have $\sum_{i=1}^{N} \varsigma_i \alpha_i \neq 0$. The quantity
\be
	\bm{\delta_\Sigma} (\vartheta) \equiv \min_{\varsigma \in \{\pm 1\}^{N}} \Ab{\sum_{i = 1}^{N} \varsigma_i \alpha_i}
\ee
is called the minimal discrepancy \cite{partition06}, and the numbers in the set $\vartheta$ are called impartitionable if, and only if $\bm{\delta_\Sigma} (\vartheta) > 0$. In other words, a set of real numbers $\vartheta$ is impartitionable, if one cannot find a partition of the set $\vartheta$ into two subsets summing to the same value.

It is easy to see that if $\{g_i\}_{i=1}^{N}$ are independent and continuously distributed random variables, then for any $\varsigma \in \{\pm 1\}^{N}$ and $t > 0$ we have $P \left( \sum_{i=1}^{N} \varsigma_i g_i t = 0 \right) = 0$, so the coupling constants are impartitionable almost surely. Using this notion in Appendix~\ref{sec:avgPeriodic} we prove the following:
\begin{prop}
	\label{prop:timeAvg}
	If the coupling constants $\{g_i\}_{i=1}^{N}$ are impartitionable, then the long time averages of $B(t)^2$ and $\Ab{\gamma(t)}^2$ are given by the formulae:
	\begin{subequations}
		\be
			\label{eq:B2TimeAvg}
			\ba
				\overline{B^2} &\equiv \lim_{T \to \infty} \frac{1}{T} \int_0^T B^2(t) dt\\
				&= \prod_{j \in mac} \left[ 1 - \frac{(2 \lambda_j - 1)^2 \sin^2 \beta_j}{2} \right],
			\ea
		\ee
		\be
			\label{eq:gamma2TimeAvg}
			\overline{\Ab{\gamma}^2} \equiv \lim_{T \to \infty} \frac{1}{T} \int_0^T \Ab{\gamma(t)}^2 dt = \prod_{j \notin Obs} \frac{1 + \zeta_j^2}{2}.
		\ee
	\end{subequations}
\end{prop}

From the Proposition~\ref{prop:timeAvg} it follows directly that for an ensemble in which the coupling constants $\{g_i\}_{i=1}^{N}$ are impartitionable almost surely, the long time averages of $B(t)^2$ and $\Ab{\gamma(t)}^2$ do not depend on the distributions of the coupling constants. 

In order to get some intuition about the Proposition~\ref{prop:timeAvg}, let us note that the time average of a product of periodic functions in~\eqref{eq:B} and~\eqref{eq:gamma2}, all with different periods almost surely (assuming a non-degenerate distribution of the coupling constants) is equal to the product of time averages of each function, if the time is long enough. On the other hand, for a particular term in the product the value of the coupling constant influences only the period, not the average value.

\section{The objectivization process of a spin register}
\label{sec:register}

We now extend the above analysis to a central system consisting of $K$ spins with the evolution given by the Hamiltonian~\eqref{eq:registerHint}. Each register spin can now interact with all of the environmental spins and the interaction strength is controlled by the coupling matrix $G = [g_{ij}]$.

As shown e.g. in~\cite{Nalezyty} the spin register model exhibits a dynamical structure with the existence of so-called Decoherence Free Subspaces (DFS). We also introduce and investigate here the notion of Orthogonalization Free Subspaces (OFS).

\subsection{The orthogonalization and decoherence factors}
\label{sec:spinRegister}

Derivation of the evolution operator from~\eqref{eq:registerHint} is straightforward:
\be
	\label{eq:USE}
	U_{S:E}(t) = \sum_{\bm{\epsilon} \in \{\pm 1\}^K } \proj{\bm\epsilon} \otimes \bigotimes_{j=1}^N U_{\bm\epsilon}^{(j)}(t),
\ee
where $\bm\epsilon\equiv (\epsilon_1,\dots,\epsilon_K)$ denotes a bit-string of length $K$ and
\be
	U_{\bm\epsilon}^{(j)}(t)\equiv \exp \left[ -\frac{i}{2} t \left( \sum_{i=1}^K\epsilon_i g_{ij} \right)\sigma_z^{(j)} \right],
\ee
cf.~\eqref{eq:Uqml}. Let us denote:
\be
	\label{eq:evolRho}
	\ba
		& \rho_{\bm{\epsilon\epsilon'}}^{(j)}(t) \equiv U_{\bm\epsilon}^{(j)}(t) \rho^{(j)}(0) U_{\bm\epsilon'}^{(j) \dagger}(t), \\
		& \rho_{\bm{\epsilon}}^{(j)}(t) \equiv \rho_{\bm{\epsilon\epsilon}}^{(j)}(t)
	\ea
\ee
cf.~\eqref{eq:varrhoI}, and similarly as in~\eqref{eq:rhoMac} we define for a given $\bm{\epsilon}$ the macrofraction state $\varrho_{\bm{\epsilon}}^{(k)}(t) \equiv \bigotimes_{j \in mac_k} \rho_{\bm{\epsilon}}^{(j)}(t)$.

Calculation of the partially traced state $\varrho_{S:fM}(t)$ as in~\eqref{reduced}, gives:
\be
	\label{eq:registerTraced}
	\ba
		\varrho_{S:fM}(t) = & \sum_{\bm{\epsilon}} \sigma_{\bm\epsilon} \proj{\bm\epsilon} \otimes \bigotimes_{j \in Obs} \rho_{\bm{\epsilon}}^{(j)}(t) \quad + \\
		& \sum_{\bm{\epsilon}\ne\bm{\epsilon'}} \sigma_{\bm{\epsilon\epsilon'}} \gamma_{\bm{\epsilon\epsilon'}}(t) \ket{\bm\epsilon} \bra{\bm\epsilon'} \otimes \bigotimes_{j \in Obs} \rho_{\bm{\epsilon\epsilon'}}^{(j)}(t).
	\ea
\ee
Notation $\bm{\epsilon} \ne \bm{\epsilon'}$ means that there exists at least one index $i$ such that $\epsilon_i \ne \epsilon'_i$. As one sees from~\eqref{eq:registerTraced}, the candidate for the pointer basis is now the product basis $\ket{\bm\epsilon}\equiv\ket{\epsilon_1,\dots,\epsilon_K}$, describing a classical $K$-bit register. Repeating the steps (\ref{eq:M})-\eqref{eq:B}, it is easy to see that the overlap functions for $k$-th macrofraction (now there are more of them than just one as before),
\be
	B_{\bm{\epsilon\epsilon'}}^{(k)}(t) \equiv B[\varrho_{\bm{\epsilon}}^{(k)}(t),\varrho_{\bm{\epsilon'}}^{(k)}(t)],
\ee
are given by the same formula~\eqref{eq:B}, but with different time frequencies, viz.:
\be
	\label{eq:Bepseps}
	B_{\bm{\epsilon\epsilon'}}^{(k)}(t) = \prod_{j \in mac_k} \sqrt{1 - (2 \lambda_j - 1)^2 \sin^2 \beta_j \sin^2(\omega^j_{\bm{\epsilon\epsilon'}} t)},
\ee
where we introduced the following frequency:
\be
	\label{eq:omegaEpsEps}
	\omega^j_{\bm{\epsilon\epsilon'}} \equiv \sum_{i=1}^K (\epsilon_i - \epsilon'_i) g_{ij}.
\ee

The decoherence factors, cf.\eqref{eq:gamma2}, are given by:
\be
	\label{gamma2}
	\ba
		\gamma_{\bm{\epsilon\epsilon'}}(t) & \equiv \prod_{j \notin Obs} \Tr \left[\exp \left( \frac{i}{2} t \left( \sum_{i=1}^K (\epsilon_i-\epsilon'_i) g_{ij} \right) \sigma_z^{(j)} \right) \varrho^{(j)}(0) \right] \\
		& = \prod_{j \notin Obs} \left[ \cos\left(\omega^j_{\bm{\epsilon\epsilon'}} t \right) + i \zeta_j \sin\left(\omega^j_{\bm{\epsilon\epsilon'}}t \right) \right].
	\ea
\ee

\subsection{The orthogonalization and decoherence free subspaces}

We say that a subspace $S \subseteq \{\pm\}^{K}$ exhibits the strong DFS property if
\be
	\bigforall_{t \in \mathbb{R}_+} \bigforall_{\bm{\epsilon},\bm{\epsilon'} \in S} \gamma_{\bm{\epsilon\epsilon'}}(t) = 1,
\ee
and a weak DFS if
\be
	\bigforall_{t \in \mathbb{R}_+} \bigforall_{\bm{\epsilon},\bm{\epsilon'} \in S} \Ab{\gamma_{\bm{\epsilon\epsilon'}}(t)} = 1.
\ee

From~\eqref{gamma2} it is easy to see that the strong DFS holds if, and only if
\be
	\label{eq:no-decoherence}
	\bigforall_{\bm{\epsilon},\bm{\epsilon'} \in S} \bigforall_{j \notin Obs} \omega^j_{\bm{\epsilon\epsilon'}} = 0,
\ee
and the weak DFS occurs if we have
\be
	\bigforall_{\bm{\epsilon},\bm{\epsilon'} \in S} \bigforall_{j \notin Obs} \left[ \omega^j_{\bm{\epsilon\epsilon'}} = 0 \vee \zeta_j = 1 \right].
\ee
The strong DFS means that the register state remains invariant under time evolution rather than being unitarily rotated inside a DFS: this is a much more desired property from experimentalist point of view, as such rotation could lead to the system being uncontrollable~\cite{Nalezyty}.

Similarly we define OFS to occur in the case when
\be
	\bigforall_{\bm{\epsilon},\bm{\epsilon'} \in S} \bigforall_{k=1, \cdots fM} B_{\bm{\epsilon\epsilon'}}^{(k)}(t) = 1,
\ee
which by~\eqref{eq:Bepseps} holds if, and only if
\be
	\label{eq:no-orth}
	\bigforall_{\bm{\epsilon},\bm{\epsilon'} \in S} \bigforall_{j \in Obs} \omega^j_{\bm{\epsilon\epsilon'}} = 0.
\ee

It is easy to see, that the only difference between the definitions of the strong DFS~\eqref{eq:no-decoherence}, and OFS~\eqref{eq:no-orth}, is the scope of one of the universal quantificators, covering all spins outside or inside the observed part, respectively. The coupling matrix can be therefore divided in the following way:
\be
	G = \left[\smash{\underbrace{
\begin{array}{ccc}
	g_{11} & \ldots & g_{1,\Ab{Obs}} \\
	\vdots & \ddots & \vdots	 \\
	g_{K,1} & \ldots & g_{K,\Ab{Obs}}
\end{array}}_{\text{observed}}}
\begin{array}{c|c}
	&\\
	&\\
	&
\end{array}
\smash{\underbrace{
\begin{array}{ccc}
	g_{1,\Ab{Obs}+1} & \ldots & g_{1,N} \\
	\vdots		 & \ddots & \vdots \\
	g_{K,\Ab{Obs}+1} & \ldots & g_{K,N} 
\end{array}}_{\text{unobserved}}}
\right].
\ee
\newline
We see that the matrix $G$ has a block structure $G=[G_1 | G_2]$ where $G_1$, $G_2$ describe the interaction with the observed and unobserved part of the environment, respectively. From~\eqref{eq:omegaEpsEps} it is obvious that for~\eqref{eq:no-decoherence} and~\eqref{eq:no-orth} to hold $\bm{\epsilon}-\bm{\epsilon'}$ has to be in the kernel of $G_2^T$ and $G_1^T$, respectively.

Now, let us deal in more details with DFSs and OFSs.

\subsection{All of $\omega^j_{\bm{\epsilon\epsilon'}} \ne 0$}

The analysis here is essentially the same as in the single central spin case studied in sec.~\ref{sec:spinSpin}, only with different time frequencies.

In particular, the short-time behavior~\eqref{shorttime} is now controlled by the coupling-averaged quantities
\be
	\E{ (\omega^j_{\bm{\epsilon\epsilon'}})^2 } = \sum_{i,i'=1}^K (\epsilon_i - \epsilon'_i) (\epsilon_{i'} - \epsilon'_{i'}) \E{g_{i j} g_{i' j}},
\ee
where we assume that the average $\E{g_{i j} g_{i' j}}$ exists and the $G$ matrix can be modeled as a random matrix.

For large enough macrofractions and long enough times (defined now w.r.t. $\omega^j_{\bm{\epsilon\epsilon'}}$), the partially traced state approaches the SBS form, meaning that the spin register has decohered in the classical register basis $\ket{\bm\epsilon}$ and the information about this register is redundantly stored in the environment.

\subsection{For some $\bm{\epsilon},\bm{\epsilon'}$ and all $j$: $\omega^j_{\bm{\epsilon\epsilon'}}=0$} 

We assume there are only two strings of bits $\bm\epsilon,\bm{\epsilon'}$ with such a property; if there are more the analysis is analogous. This is the OFS and strong DFS case. From~\eqref{gamma2} one sees that the coherence between the states $\ket{\bm\epsilon}$ and $\ket{\bm{\epsilon'}}$ is preserved by the evolution. We also have:
\be
	\label{U=U}
	U_{\bm\epsilon}^{(j)}(t)=U_{\bm{\epsilon'}}^{(j)}(t)
\ee
for all environments $j$. Thus $U_{S:E}(t)$, cf.~\eqref{eq:USE}, is given by:
\be
	 \Pi_{\bm{\epsilon\epsilon'}} \otimes \bigotimes_{j=1}^N U_{\bm\epsilon}^{(j)}(t) + 
		\sum_{\bm{\epsilon''} \ne \bm{\epsilon},\bm{\epsilon'}} \proj{\bm{\epsilon''}} \otimes \bigotimes_{j=1}^N U_{\bm{\epsilon''}}^{(j)}(t),
\ee
where $\Pi_{\bm{\epsilon\epsilon'}} \equiv \proj{\bm\epsilon} + \proj{\bm{\epsilon'}}$ is the projector on the register subspace spanned by $\{\ket{\bm\epsilon},\ket{\bm{\epsilon'}}\}$. In particular $B_{\bm{\epsilon\epsilon'}}(t)=\gamma_{\bm{\epsilon\epsilon'}}(t)=1$.

If there are no other DFSs and the conditions for formation of the broadcast state are met apart from the pair $\bm{\epsilon}$, $\bm{\epsilon'}$, i.e. all the decoherence and orthogonalization factors apart from $B_{\bm{\epsilon\epsilon'}}$ and $\gamma_{\bm{\epsilon\epsilon'}}$ disappear, the partially traced state approaches what we call a coarse-grained SBS:
\be
	\label{coarseSBS}
	\ba	
	\varrho_{S:fM}(t) &= 
		\Pi_{\bm{\epsilon\epsilon'}} \varrho_S(0) \Pi_{\bm{\epsilon\epsilon'}}
		\otimes
		\bigotimes_{j \in Obs} \rho_{\bm{\epsilon}}^{(j)}(t) \\
		&+
		\sum_{\bm{\epsilon''} \ne \bm{\epsilon}, \bm{\epsilon'} } \sigma_{\bm{\epsilon''}} \proj{\bm{\epsilon''}} \otimes
		\bigotimes_{j \in Obs} \rho_{\bm{\epsilon''}}^{(j)}(t).	
	\ea
\ee
Information that leaked into the environment about the register's state $\ket{\bm\epsilon}$ is not complete, viz. it is impossible to tell if the register is in the state $\ket{\bm\epsilon}$ or $\ket{\bm{\epsilon'}}$ by observing the environment, and this holds no matter how big the macrofractions are. The information is simply not in the environment. Moreover, the $\ket{\bm\epsilon}$ and $\ket{\bm{\epsilon'}}$ block of the initial state $\varrho_S(0)$ is fully preserved by the dynamics in this case.

\subsection{For some $\bm{\epsilon},\bm{\epsilon'}$: $\omega^j_{\bm{\epsilon\epsilon'}}=0$ for all $j \in Obs$, and $\omega^j_{\bm{\epsilon\epsilon'}} \ne 0$ for all $j \notin Obs$}
\label{sec:noOrthDec}

From~\eqref{eq:Bepseps} and~\eqref{gamma2} it follows that the decoherence takes place but the orthogonalization does not. This is a relatively common situation in real life, when the environment is unable to store faithfully information about the decoherening system (e.g. due to too high intrinsic noise as compared to the interaction strength). 

The property~\eqref{U=U} holds again here for all $j$ in all observed macrofractions, which by~\eqref{eq:evolRho} implies that $\rho_{\bm{\epsilon}}^{(j)}(t) = \rho_{\bm{\epsilon'}}^{(j)}(t)$ (as mentioned we describe here idealized cases for the sake of simplicity). Thus, the resulting asymptotic state is similar to~\eqref{coarseSBS} but with destroyed coherences:
\be
	\ba
		\varrho_{S:fM}(t) &= 
		\left( \sigma_{\bm\epsilon} \proj{\bm\epsilon} + \sigma_{\bm{\epsilon'}} \proj{\bm{\epsilon'}} \right) \otimes
		\bigotimes_{j \in Obs} \rho_{\bm{\epsilon}}^{(j)}(t) \\
		&+
		\sum_{\bm{\epsilon''} \ne \bm{\epsilon},\bm{\epsilon'}} \sigma_{\bm{\epsilon''}} \proj{\bm{\epsilon''}} \otimes
		\bigotimes_{j \in Obs} \rho_{\bm{\epsilon''}}^{(j)}(t).
		%
	\ea
\ee
Observing the string of bits $\bm{\epsilon}$ in the environment only tells us that the system is with probability $\frac{\sigma_{\bm{\epsilon}}}{\sigma_{\bm{\epsilon}}+\sigma_{\bm{\epsilon'}}}$ in the state $\ket{\bm{\epsilon}}$ and with probability $\frac{\sigma_{\bm{\epsilon'}}}{\sigma_{\bm{\epsilon}}+\sigma_{\bm{\epsilon'}}}$ in the state $\ket{\bm{\epsilon'}}$.

\subsection{For some $\bm{\epsilon},\bm{\epsilon'}$: $\omega^j_{\bm{\epsilon\epsilon'}} \ne 0$ for all $j \in Obs$, and $\omega^j_{\bm{\epsilon\epsilon'}}=0$ for all $j \notin Obs$}
\label{sec:orthNoDec}

This is a reversed situation to the one above: The decoherence does not take place, cf.~\eqref{gamma2}, but the orthogonalization does, cf.~\eqref{eq:Bepseps}.

This situation is quite peculiar and only possible because orthogonalization and decoherence are driven by different parts of the environment: The observed and the unobserved, respectively. Otherwise, one can prove that for the same portion of the environment it always holds $\Ab{\gamma(t)} \leq B(t)$ \cite{ineq}. The asymptotic state is in this case given by:
\be
	\label{strange}
	\ba
		\varrho_{S:fM}(t) &=
		\sum_{\bm\tilde\epsilon}\sigma_{\bm\tilde\epsilon} \proj{\bm\tilde\epsilon} \otimes
		\bigotimes_{j \in Obs} \rho_{\bm{\epsilon}}^{(j)}(t)
		\\
		&+
		\sigma_{\bm{\epsilon\epsilon'}} \gamma_{\bm{\epsilon\epsilon'}}(t) \ket{\bm\epsilon} \bra{\bm{\epsilon'}} \otimes
		\bigotimes_{j \in Obs} \rho_{\bm{\epsilon\epsilon'}}^{(j)}(t)
		+\text{h.c.}
		%
	\ea
\ee
where, for each $j$, $\varrho_{\bm{\tilde\epsilon}}^{(j)}$ are fully distinguishable for all $\bm{\tilde\epsilon}$, including $\varrho_{\bm{\epsilon}}^{(j)}$ and $\varrho_{\bm{\epsilon'}}^{(j)}$.

The state~\eqref{strange} possesses what one may call a genuine multipartite coherences (which can include entanglement). Indeed, let us trace out one of the observed macrofractions, $mac$. This will produce an extra decoherence factor,
\be
	\gamma_{\bm{\epsilon\epsilon'}}^{(k)}(t) \equiv \prod_{j \in mac_k} \Tr \left[ U_{\bm\epsilon}^{(j)} \rho^{(j)}(0) U_{\bm\epsilon'}^{(j) \dagger} \right].
\ee
By the quoted property $\Ab{\gamma^{(k)}_{\bm{\epsilon\epsilon'}}(t)} \leq B^{(k)}_{\bm{\epsilon\epsilon'}}(t)$, so the assumed vanishing of the latter implies destruction of the coherences. Thus, tracing out a single macrofraction for the asymptotic state destroys all the remaining coherences and brings the state to an SBS form:
\be
	\Tr_{mac} \varrho_{S:fM} = 
	\sum_{\bm{\tilde\epsilon}} \sigma_{\bm{\tilde\epsilon}} \proj{\bm\tilde\epsilon} \otimes \varrho_{\bm{\tilde\epsilon}}^{(1)} \otimes \cdots \xcancel{\varrho_{\bm{\tilde\epsilon}}^{(k)}} \cdots \otimes \varrho_{\bm{\tilde\epsilon}}^{(fM)}.
\ee
We postpone a further investigation of those states, especially in a relation to crypthographic protocols, to a subsequent publication.

We provide theoretical examples of setups in which the above cases occur below in sec.~\ref{sec:DecOrthFreeExamples}.

\section{Examples of decoherence and orthogonalization for spin registers}
\label{sec:DecOrthFreeExamples}

Now, as an illustration, we consider several specific choices of coupling constants $g_{ij}$ such that register spins interact non-trivially with environmental ones.

\subsection{Collective decoherence}

\begin{figure}[hbpt!]
	\centering
 \def\svgwidth{0.7 \columnwidth}
	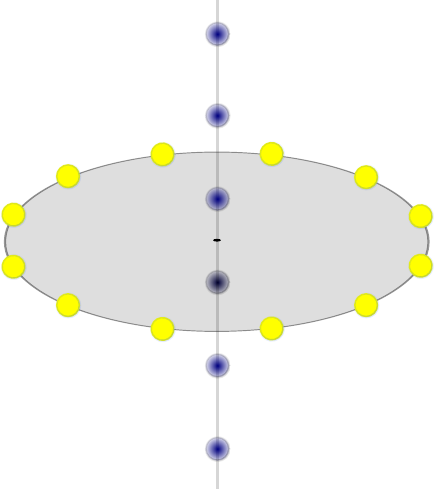
 \caption{(Color online) The geometry of the composite system leading to the collective decoherence. The register spins (yellow) are placed on a circle through the center of which passes a straight line on which the environmental spins are located (blue). The position of the environmental spins on the line is symmetrical with respect to the circle.}
	\label{fig:geometria1}
\end{figure}

Here the coefficients $g_{ij}$ are $i$ independent, which leads to exceptionally rich family of DFSs. If the coupling coefficients depend solely on the distance between the register and environmental spin, then such choice can be obtained by placing the system as shown in Fig.~\ref{fig:geometria1}. This setup can be achieved e.g. in crystalline solid with help of a scanning tunneling microscope.

Obviously, by~\eqref{eq:no-decoherence}, if the number of '+1' entries is the same for both $\vec{\bm{\epsilon}}$ and $\vec{\bm{\epsilon'}}$, then no broadcasting occurs, as no information about the system is transferred into the environment whatsoever. Thus both such register states belong to the same DFS.

\subsection{Cylindrical symmetry}

\begin{figure}[hbpt!]
	\centering
 \def\svgwidth{0.7 \columnwidth}
	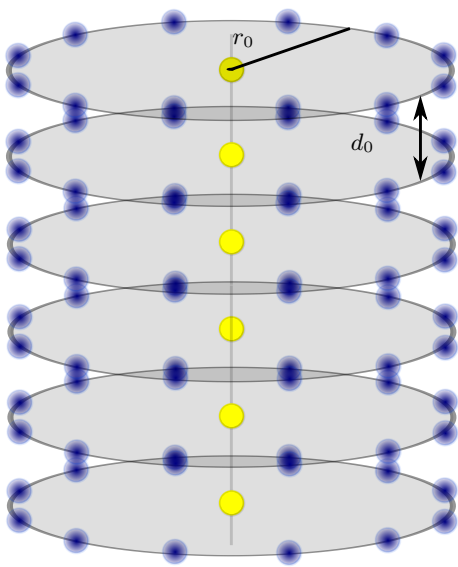
 \caption{(Color online) A different geometry with a linear register (yellow) and a collection of environmental spins (blue) located on $M$ circles of $L$ spins. The radius of each circle is $r_0$, while the distance between neighboring circles is $d_0$.}
	\label{fig:geometria2}
\end{figure}

Let us now consider a generalization of the above example with the register and environmental spins organized as shown in Fig.~\ref{fig:geometria2}. The register spins are located on a straight line and the environment is composed of a collection of $K$ circles, each containing $L$ spins. The geometry is chosen such that $k$-th register spin is coplanar with $k$-th circle.

The locations of environmental spins are most easily expressed by the number of the circle to which it belongs and its location on the circle. Due to the geometry of the composite system, the coupling constants are independent of the latter. We assume that they depend on the distance between $i$-th register spin and $l$-th spin on $m$-th circle as
\be
	g_{iml} = g_{im} = \frac{g_0}{r_{im}^3} \approx \frac{g_0}{r_0^3} \left[ 1-\frac{3}{2} (i-m)^2 \frac{d_0^2}{r_0^2} \right],
\ee
where for the approximation we used the Taylor expansion and assumed that $r_0 \gg M d_0$. The meaning of the symbols is explained conceptually in Fig.~\ref{fig:geometria2}.

The frequencies $\omega^{(ml)}_{\bm{\epsilon\epsilon'}} = \omega^{m}_{\bm{\epsilon\epsilon'}}$
by~\eqref{eq:omegaEpsEps} can be found to be
\be
	\ba
		\omega^{m}_{\bm{\epsilon\epsilon'}} & = \sum_{i = 1}^{K} \left( \epsilon_i - \epsilon'_i \right) g_{im} \\
		& = -\frac{3}{2} \frac{g_0 d_0^2}{r_0^5} \sum_{i = 1}^{K} \left( \epsilon_i - \epsilon'_i \right) \left[ -\frac{2}{3} \frac{r_0^2}{d_0^2} + (i - m)^2 \right].
	\ea
\ee
Now, as discussed before, the no-decoherence criterion demands that for each $m$ the frequency should be equal to 0. Observe that $\omega^{m}_{\bm{\epsilon\epsilon'}}$ can be regarded as a quadratic function of $m$ of the form $am^2 + bm +c$, where:
\begin{gather}
	\ba
		a &= \sum_{i=1}^{K} (\epsilon_i - \epsilon_i'),	\\
		b &= \sum_{i=1}^{K} (\epsilon_i - \epsilon_i') i,	\\
		c &= \sum_{i=1}^{K} (\epsilon_i - \epsilon_i') (i^2 - \frac{2}{3} \cdot \frac{r_0^2}{d_0^2}).
	\ea
\end{gather}
For~\eqref{eq:no-decoherence} to be met one must have $a=b=c=0$, or equivalently:
\begin{gather}
	\label{eq:model2_no-dec}
	\ba
		\sum_{i=1}^{K} (\epsilon_i - \epsilon_i') &= 0,	\\
		\sum_{i=1}^{K} (\epsilon_i - \epsilon_i') i &= 0,	\\
		\sum_{i=1}^{K} (\epsilon_i - \epsilon_i') i^2 &= 0.
	\ea
\end{gather}
Note that the difference $\epsilon_i - \epsilon_i'$ can take values either $0$ or $\pm 2$.

\begin{obs}
	\label{lem:subsets}
	There exist an infinite number of the systems of equations~\eqref{eq:model2_no-dec} with non-trivial solutions, i.e. there exist an infinite number setups of this kind with DFS.
	\begin{proof}
		For a given $K \in \mathbb{N}_{+}$ let us denote $[K] = \{1, \cdots, K\}$.
		
		The number of possible sizes of subsets of $[K]$ is $K + 1$, the number of possible sums of their elements is at most of order $K^2$, and the number of possible sums of squares of their elements is at most of the order $K^3$. So, there is at most $(K+1) \cdot K^2 \cdot K^3 = (K+1) \cdot K^5$ different values of the triple of size, sum of elements and sum of squares of elements for subsets of $[K]$.
		
		On the other hand there exist $2^K$ different subsets of the set $[K]$.	Thus, when $2^K > (K+1) \cdot K^5$, there must exist at least two subsets with equal number of elements, their sum and sum of squares.
	\end{proof}
\end{obs}
An example of pairs of sets occurring in the proof of Lemma~\ref{lem:subsets} is $\{1,5,6\}$ and $\{2,3,7\}$ for $K=7$; these sets defines which elements of vectors $\vec{\bm{\epsilon}}$ and $\vec{\bm{\epsilon'}}$, respectively, have $+1$ value, with $-1$ at other places.

\subsection{Decoherence--free and orthogonalization--free processes}

\begin{figure}[hbpt!]
	\centering
 \def\svgwidth{ \columnwidth}
	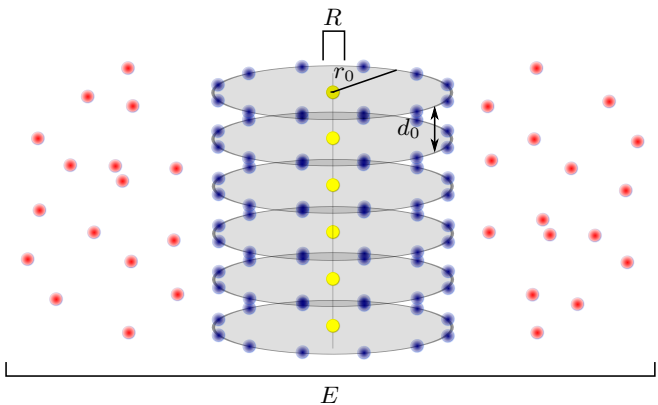
 \caption{(Color online) The geometry of the extended register model. The environment $E$ now consists not only of the cylindrically placed spins (blue) but also of additional spins placed randomly in the space (red).}
	\label{fig:clean}
\end{figure}

We apply the result of sec.~\ref{sec:register} to the extended linear register model, with the geometry given in Fig.~\ref{fig:clean} and show examples of setups leading to the cases with decoherence and no orthogonalization (sec.~\ref{sec:noOrthDec}) and with orthogonalization and no decoherence (sec.~\ref{sec:orthNoDec}).

The environment consists of two parts: the first contains a collection of cylindrically placed spins (blue) as in the original model, while the spins of the second part (red) are randomly distributed in space. Let us now analyze the two cases separately.

\subsubsection{Decoherence but no orthogonalization}

\begin{figure}[hbpt]
	\centering
 \def\svgwidth{ \columnwidth}
	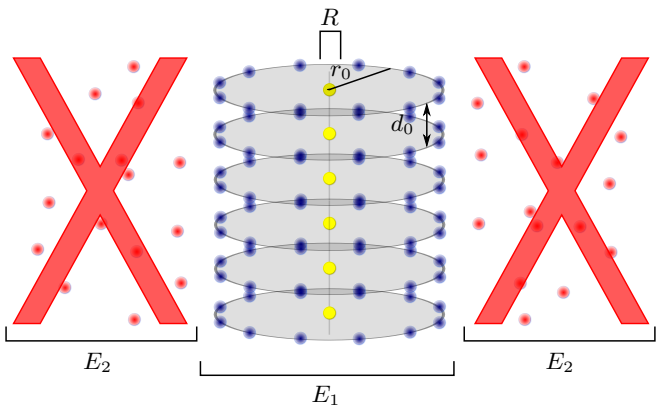
 \caption{(Color online) A geometry that allows for case with decoherence but no orthogonalization to occur. Frequencies $\omega^k_{\bm{\epsilon\epsilon'}}$ can vanish for spins in $E_1$, but are generally non-zero for the randomly located ones from $E_2$.}
	\label{fig:a}
\end{figure}

Let the observable environment macrofraction ($E_1$) be the `blue' spins (see Fig.~\ref{fig:a}), while we discard the 'random' ones. As it was shown before, within this model it is possible to choose two states, $\ket{\bm{\epsilon}}$ and $\ket{\bm{\epsilon'}}$, for which the frequencies corresponding to $E_1$ vanish, and therefore no orthogonalization occurs.

However, location of 'red' spins means that $G_2$ is a random matrix, thus the decoherence factor computed from it is a quasi-periodic function with generally a very big recurrence time, so that the system is effectively a decohering one.

\subsubsection{Orthogonalization but no decoherence}

\begin{figure}[hbpt!]
	\centering
	\def\svgwidth{ \columnwidth}
	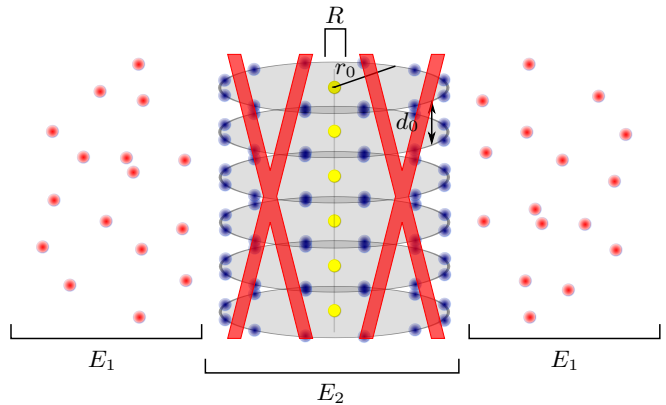
	\caption{(Color online) A geometry that allows for case 4 to occur (orthogonalization but no decoherence). Here the situation is reversed as compared with Fig.~\ref{fig:a}}
	\label{fig:b}
\end{figure}

This situation is actually the opposite to the previously considered one with the roles of $E_1$ and $E_2$ exchanged, see Fig.~\ref{fig:b}. Now, if the states $\ket{\bm{\epsilon}}$ and $\ket{\bm{\epsilon'}}$ are chosen properly, they do not decohere, but the orthogonalization still takes place as the frequencies pertaining to the observable macrofraction do not vanish in general.

\subsection{Small overlap of interactions}

Suppose that the coupling coefficients in $G$ are given by the following function:
\be
	\label{eq:small_over_g}
	g_{ij} = f_{\mu_i,\sigma} (j),
\ee
where $f_{\mu,\sigma}:\ \mathbb{R}\rightarrow \mathbb{R}$ is a $C^0$ function describing a normalized 'saw pulse' centered around $\mu$ and of width $\sigma$, given by:
\be
	f_{\mu,\sigma}(x) =
	\begin{cases}
		0 & \text{for } x \in (-\infty, \mu - \frac{\sigma}{2}), \\
		\frac{2}{\sigma} x + (1-\frac{2\mu}{\sigma}) & \text{for } x \in [\mu-\frac{\sigma}{2} , \mu), \\
		-\frac{2}{\sigma} x + (1+\frac{2\mu}{\sigma}) & \text{for } x \in [\mu , \mu+\frac{\sigma}{2}), \\
		0 & \text{for } x \in [\mu + \frac{\sigma}{2}, \infty).
	\end{cases}
\ee

\begin{figure}[hbpt!]
	\centering
	\includegraphics[width=\columnwidth]{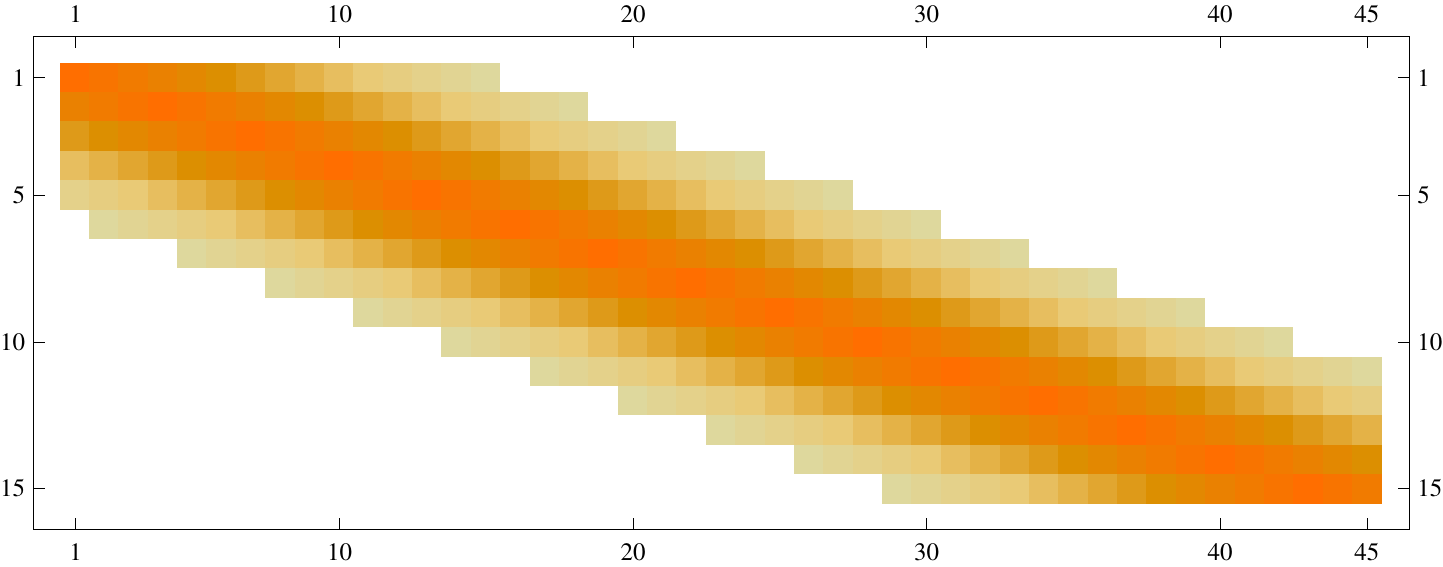}
	\caption{(Color online) The structure of the coupling matrix $G$ with the values chosen as in~\eqref{eq:small_over_g} with $\sigma=30$. The vertical and horizontal axes correspond to the row and column of $G$, respectively. The values of the coupling coefficients are expressed using coloring technique (higher intensity represents higher value).}
	\label{fig:so1}
\end{figure}

Furthermore, we assume that 
\be
	\mu_i = \frac{N}{K}(i-1)+1, i = 1, \dots, K.
\ee
Recall that $N$ is the number of spins in environment, and $K$ in the register. Such a choice states that entries of successive rows of $G$ are given by such 'saw pulses' with equidistant maxima and the same width, see Fig.~\ref{fig:so1}.

\begin{figure}[hbpt!]
	\centering
	\includegraphics[width=\columnwidth]{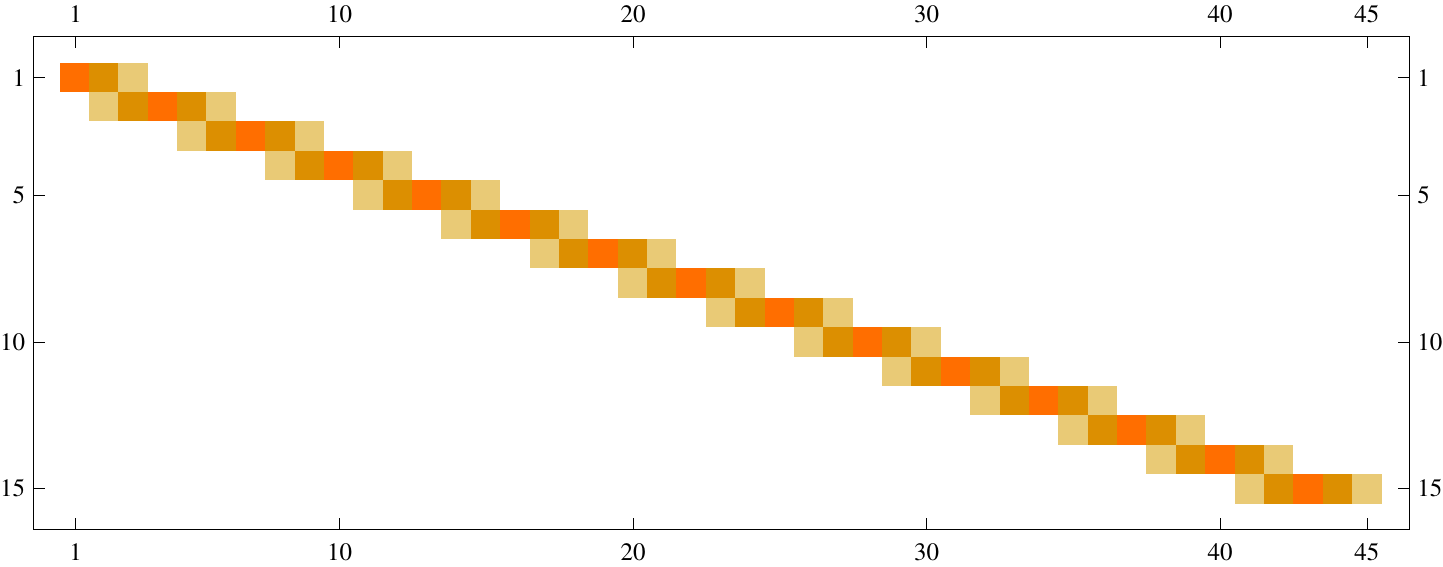}
	\caption{(Color online) An example of a coupling matrix $G$ leading to the small overlap case. Note that the overlap of an arbitrary $i$-th row with its neighboring ones is such that they vanish in the vicinity of its maximum.}
	\label{fig:so2}
\end{figure}

Moreover we assume that the overlap between any two consecutive rows of $G$ is small, i.e.:
\be
	\label{eq:small_overlap}
	\frac{\sigma}{2}<\frac{N}{K},
\ee
which simply means that the separation between the maxima of two successive rows is greater than the width $\sigma$ of the function $f_{\mu,\sigma}$ (see Fig.~\ref{fig:so2}).

From~\eqref{eq:no-decoherence} it follows that given such assumptions the strong decoherence condition cannot be fulfilled. We see that the coupling coefficient matrix $G$ is of full rank, and therefore the decoherence occurs. Indeed, consider $i$-th row of $G$: if $G$ was not of full rank, then it would be possible to express this row as a linear combination of the remaining ones. But due to the assumption	\eqref{eq:small_overlap} this is impossible, as in the region close to the peak of $i$-th row the entries of all other ones are equal to zero.

\section{Conclusions}

We studied in detail decoherence process of a qubit register, coupled to a spin environment through a $ZZ$ interaction. Our main interest was in the information transfer from the spin register to the environment. This required a departure from the standard approach as not all of the environment could be traced out.

Following our earlier research, the main object of the study was, so called, partially traced state, obtained from the full system-environment state by tracing out only a fraction of the environment. In particular, we were interested if the partially traced state approaches what we call a Spectrum Broadcast Structure. It implies a certain objectivization of a (decohered) state of the register: A classical bit-string, labelling decohered states of the register, is present in the environment in many copies and can be read out without any disturbance (on average).
 
Exploiting certain properties of quasi-periodic functions, we formulated and studied conditions when SBS's are formed asymptotically. Due to the presence of decoherence free subspaces and, what we call orthogonalization free subspaces, possible structures that can appear are much richer than in the case of a single central spin. In particular, we reported a new kind of SBS, where some of the coherences are preserved but still some information is objective. We also presented a series of theoretical examples, illustrating how different forms of the coupling matrix can lead to different patters of information proliferation. The next possible step would be to design a concrete experimental proposal around the presented examples.

\section{Acknowledgments}

The work was made possible through the support of grant from the John Templeton Foundation. The opinions expressed in this publication are those of the authors and do not necessarily reflect the views of the John Templeton Foundation. It was also supported by a National Science Centre (NCN) grant 2014/14/E/ST2/00020 and DS Programs of the Faculty of Electronics, Telecommunications and Informatics, Gda\'nsk University of Technology.

\begin{appendix}
\section{Orthogonalization and decoherence processes in large environments}\label{largev}

Let us concentrate on a particular macrofraction, $mac = mac_k$ for some $k$, $B(t) \equiv B^{(k)}(t)$, $N_{mac} \equiv \Ab{mac}$. Let us also define $N_{dis} \equiv N - \Ab{Obs}$, the number of the discarded, unobserved spins. From~\eqref{eq:B} and~\eqref{eq:gamma2} we can directly obtain:
\begin{subequations}
	\label{eqs:logBgamma}
	\be
		\log B(t) \leq - \frac{1}{2} \sum_{j \in mac} \kappa_j^t,
	\ee
	\be
		\log \Ab{\gamma(t)}^2 \leq - \sum_{j \notin Obs} \chi_j^t,
	\ee
\end{subequations}
with
\begin{subequations}
	\label{eqs:kappaChi}
	\be
		\ba
			0 \leq \kappa_j^t & \equiv (2 \lambda_j - 1)^2 \sin^2 \beta_j \sin^2(g_j t) \leq \\
			& \leq - \log \left[ 1 - (2 \lambda_j - 1)^2 \sin^2 \beta_j \sin^2(g_j t) \right],
		\ea
	\ee
	\be
		\ba
			0 \leq \chi_j^t & \equiv \sin^2 (g_j t) \left(1 - (2 \lambda_j - 1)^2 \cos^2 \beta_j \right) \leq \\
			& \leq - \log \Ab{ \cos \left(g_j t\right) + i \zeta_j \sin \left(g_j t\right) }^2.
		\ea
	\ee
\end{subequations}

We can also calculate that for the probability distributions mentioned in sec.~\ref{sec:asympt} of parameters and $t > 0$ we have
\begin{subequations}
	\label{eqs:kappaChiBounds}
	\be
		\ba
			\E{\kappa_j^t} &= \E{(2 \lambda_j - 1)^2} \E{ \sin^2 \beta_j} \E{\sin^2 (g_j t)} \\
			&= \frac{2}{5} \E{\sin^2 (g_j t)} > 0,
		\ea
	\ee
	\be
		\ba
			\E{\chi_j^t} &= \left(1 - \E{(2 \lambda_j - 1)^2} \E{\cos^2 \beta_j} \right) \E{\sin^2 (g_j t)} \\
			&= \frac{4}{5} \E{\sin^2 (g_j t)} > 0.
		\ea
	\ee
\end{subequations}

We now prove Proposition~\ref{prop:LLN}:



Let us recall that a sequence of random variables, $(X_N)_N$, satisfies LLN if for $S_N \equiv X_1 + \cdots X_N$ we have:
\be
	\label{eq:LLNlim}
	\frac{1}{N} S_N \xrightarrow[N \to \infty]{} \frac{1}{N} \E{S_N},
\ee
where the convergence is in probability. One can show~\cite{LLN} that LLN holds if\footnote{In fact the stated condition is sufficient even for the Strong Law of Large Numbers to hold, where the convergence is almost sure.} $(X_N)_N$ are independent and identically distributed (i.i.d.) and $\E{\Ab{X_1}} < \infty$. Then~\eqref{eq:LLNlim} means:
\be
	\frac{1}{N} S_N \xrightarrow[N \to \infty]{} \E{X_1}
\ee
in probability, i.e.:
\be
	\label{eq:LLN}
	\bigforall_{\epsilon_1, \epsilon_2 > 0} \bigexists_{N_0} \bigforall_{N \geq N_0} P \left( \Ab{\frac{S_N}{N} - \E{X_1}} \leq \epsilon_1 \right) \geq 1 - \epsilon_2.
\ee

From~\eqref{eqs:kappaChi} it follows that $\E{\Ab{\kappa_j}}$ and $\E{\Ab{\chi_j}}$ are finite for any probability distributions, and if we assume they are i.i.d., we can apply LLN for the right hand sides of~\eqref{eqs:logBgamma}.

Let us fix some $\epsilon > 0$ and $t > 0$. From LLN it follows that there exist $N_{0,mac}^t$ and $N_{0,dis}^t$ such that for all $N_{mac} \geq N_{0,mac}^t$ and $N_{dis} \geq N_{0,dis}^t$ with probability at least $1 - \epsilon$:
\begin{subequations}
	\label{eqs:LLNnorms}
	\be
		\label{eq:LLNnormKappa}
		\Ab{ \left( \sum_{j \in mac} \kappa_j^t \right) - N_{mac} \E{\kappa_1^t} } \leq \epsilon N_{mac} \E{\kappa_1^t},
	\ee
	\be
		\label{eq:LLNnormChi}
		\Ab{ \left( \sum_{j \notin Obs} \chi_j^t \right) - N_{dis} \E{\chi_1^t} } \leq \epsilon N_{dis} \E{\chi_1^t}.
	\ee
\end{subequations}
To see this, in~\eqref{eq:LLN} we take $\epsilon_2 = \epsilon$, and $\epsilon_1 = \epsilon \E{\kappa_1^t}$ for~\eqref{eq:LLNnormKappa}, and $\epsilon_1 = \epsilon \E{\chi_1^t}$ for~\eqref{eq:LLNnormChi}. Thus from~\eqref{eqs:logBgamma} and~\eqref{eqs:LLNnorms} we get:
\begin{subequations}
	\label{eqs:BgammaExpBounds}
	\be
		B(t) \leq \exp \left[ - \frac{1}{2} \sum_{j \in mac} \kappa_j^t \right] \leq \exp \left[ -\frac{1 - \epsilon}{2} N_{mac} \E{\kappa_1^t} \right],
	\ee
	\be
		\Ab{\gamma(t)}^2 \leq \exp \left[ -\sum_{j \notin Obs} \chi_j^t \right] \leq \exp \left[ - (1 - \epsilon) N_{dis} \E{\chi_1^t} \right].
	\ee
\end{subequations}
If we take $N_{mac}$ and $N_{dis}$ satisfying the conditions:
\begin{subequations}
	\label{eqs:NmacNbounds}
	\be
		N_{mac} \geq \max \left[ N_{0,mac}^t, \frac{2 \log \frac{1}{\epsilon}}{(1 - \epsilon) \E{\kappa_1^t}} \right],
	\ee
	\be
		N_{dis} \geq \max \left[ N_{0,dis}^t, \frac{\log \frac{1}{\epsilon}}{(1 - \epsilon) \E{\chi_1^t}} \right],
	\ee
\end{subequations}
we get~\eqref{eq:propLLN}.



\section{Uniform distribution of the coupling constants}\label{uniform}

Let us consider the case when $g_j$ are i.i.d. with the uniform measure on $[0,1]$. We have:
\be
	\E{\sin^2 (g_j t)} = \frac{1}{2} - \frac{\sin (2t)}{4t} \geq \frac{1}{2} - \frac{1}{4t}.
\ee
For this distribution we can thus give a more specific bounds than~\eqref{eqs:kappaChiBounds}, viz.:
\begin{subequations}
	\label{eqs:uniformBounds}
	\be
		\E{\kappa_j^t} > \max \left[ \frac{2}{5} \left( \frac{1}{2} - \frac{1}{4t} \right) , 0 \right],
	\ee
	\be
		\E{\chi_j^t} > \max \left[ \frac{4}{5} \left( \frac{1}{2} - \frac{1}{4t}\right), 0 \right].
	\ee
\end{subequations}

Note that for $t \ll 1$ using the Taylor expansion we get $\E{\sin^2 (g_j t)} \approx \frac{t^2}{3}$.

The following analytical studies are performed in the short- and long time regimes.

\subsection{The short time behaviour}\label{st}

Let us assume that $g_jt \ll 1$ for all relevant $g_j$. Then from~\eqref{eqs:kappaChiBounds} and the Taylor expansion we get:
\begin{subequations}
	\label{eqs:kappaChiBoundsSmallT}
	\be
		\E{\kappa_j^t} = \frac{2}{5} \E{\sin^2 (g_j t)} \approx \frac{2}{5} \E{g_j^2 t^2} = \frac{2}{5} \overline{g^2} t^2,
	\ee
	\be
		\E{\chi_j^t} = \frac{4}{5} \E{\sin^2 (g_j t)} \approx \frac{4}{5} \E{g_j^2 t^2} = \frac{4}{5} \overline{g^2} t^2.
	\ee
\end{subequations}
From~\eqref{eqs:NmacNbounds} we can infer that when we want to assure that $B(t_B) \leq \epsilon$ we take\footnote{We use here the Puiseux expansion $\frac{\log \epsilon}{1 - \epsilon} = \log \epsilon + \epsilon \log \epsilon + O(\epsilon^2)$ and $\lim_{\epsilon \to 0_{+}} \epsilon \log \epsilon = 0$. This shows the importance of a careful choice of the value of $\epsilon_1$ in~\eqref{eq:LLN} in derivation of~\eqref{eqs:LLNnorms}.}
\be
	\E{\kappa_1^{t_B}} \geq \frac{2 \log \frac{1}{\epsilon}}{(1 - \epsilon) N_{mac}} \approx \frac{2 \log \frac{1}{\epsilon}}{N_{mac}},
\ee
and similarly for $\Ab{\gamma(t_D)}^2 \leq \epsilon$ we need
\be
	\E{\chi_1^{t_D}} \geq \frac{\log \frac{1}{\epsilon}}{(1 - \epsilon) N_{dis}} \approx \frac{\log \frac{1}{\epsilon}}{N_{dis}}.
\ee
Using~\eqref{eqs:kappaChiBoundsSmallT} we get the following time scales of orthogonalization and decoherence processes:
\begin{subequations}
	\be
		t_B \approx \sqrt{ \frac{5 \log \frac{1}{\epsilon}}{\overline{g^2} N_{mac}} },
	\ee
	\be
		t_D \approx \sqrt{ \frac{5 \log \frac{1}{\epsilon}}{4 \overline{g^2} N} }.
	\ee
\end{subequations}

From~\eqref{eqs:kappaChiBoundsSmallT} one directly obtains that in this time regime the decay of $B(t)$ and $\gamma(t)$ is upper bounded by the following functions, cf.~\eqref{eqs:BgammaExpBounds}:
\begin{subequations}
	\label{eqs:BgammaExpBoundsUniformShortT}
	\be
		B(t) \leq \exp \left[-\frac{1}{5}N_{mac} \overline{g^2}t^2 \right],
	\ee
	\be
		\Ab{\gamma(t)}^2 \leq \exp \left[-\frac{4}{5} N_{dis} \overline{g^2}t^2\right].
	\ee
\end{subequations}

\subsection{The long time behavior}\label{lt}
Let us now assume $g_jt\gg1$ for all relevant $g_j$. From~\eqref{eqs:kappaChiBounds} and~\eqref{eqs:BgammaExpBounds} we see that for $N_{mac}$ and $N_{dis}$ large enough the following bounds hold:
\begin{subequations}
	\be
		B(t) \leq \exp \left[-\frac{1}{5} N_{mac} \E{\sin^2 (g_j t)} \right],
	\ee
	\be
		\Ab{\gamma(t)}^2 \leq \exp \left[- \frac{2}{5} N_{dis} \E{\sin^2 (g_j t)} \right].
	\ee
\end{subequations}
Taking the limit of large $t$ in~\eqref{eqs:uniformBounds} we get~\eqref{eqs:BgammaExpBoundsUniformLargeT} for the uniform distribution of $g_j$. The inequalities hold for large $N_{mac}$ and $N_{dis}$ with arbitrarily high probability as they are derived using LLN, see Proposition~\ref{prop:LLN}.


\section{Average values of some specific quasi-periodic functions}
\label{sec:avgPeriodic}

Here we prove Proposition~\ref{prop:timeAvg}. We start with the following:
\begin{lemma}
	\label{lem:cosProds}
	If real number $\{\alpha_i\}_{i=1}^{N}$ are impartitionable, then:
	\begin{enumerate}
		\item
			\be
				\label{eq:cosProd}
				\prod_{i = 1}^{N} \cos \alpha_i = \frac{1}{2^N} \sum_{\upsilon \in \{\pm 1\}^{N}} \cos \left( \sum_{i=1}^{N} \upsilon_i \alpha_i \right),
			\ee
		\item
			\be
				\label{eq:cosProdMA}
				\lim_{T \to \infty} \frac{1}{T} \int_0^T \left[ \prod_{i = 1}^{N} \cos (\alpha_i t) \right] dt = 0.
			\ee
	\end{enumerate}
	
	\begin{proof}
		\begin{enumerate}
			\item For $N=1$ we obviously have
				\be
					\cos \alpha_1 = \frac{1}{2} \sum_{\upsilon \in \{\pm\}} \cos(\upsilon \alpha_1).
				\ee
				Let us assume that~\eqref{eq:cosProd} holds for some $N$, and let $\{\alpha_i\}_{i=1}^{N+1}$ be impartitionable. Then:
				\be
					\ba
						\prod_{i = 1}^{N+1} & \cos \alpha_i = \frac{1}{2^N} \sum_{\upsilon \in \{\pm 1\}^{N}} \cos \left( \sum_{i=1}^{N} \upsilon_i \alpha_i \right) \cos \alpha_{N+1} \\
						=& \frac{1}{2^N} \sum_{\upsilon \in \{\pm 1\}^{N}} \frac{1}{2} \left[ \cos \left(\sum_{i=1}^{N} \upsilon_i \alpha_i + \alpha_{N+1} \right) + \right. \\
						& \left. \cos \left(\sum_{i=1}^{N} \upsilon_i \alpha_i - \alpha_{N+1} \right)\right] \\
						=& \frac{1}{2^{N+1}} \sum_{\upsilon \in \{\pm 1\}^{N+1}} \cos \left( \sum_{i=1}^{N+1} \upsilon_i \alpha_i \right).
					\ea
				\ee
			\item First, let us note that for any $\alpha$ and $T > 0$:
				\be
					\label{eq:intTbound}
					\int_0^T \cos (\alpha t) dt \leq \int_0^{\frac{\pi}{\alpha}} \cos (\alpha t) dt = \frac{2}{\alpha},
				\ee
				and similarly $-\frac{2}{\alpha} \leq \int_0^T \cos (\alpha t) dt$. It is easy to see that if the numbers in $\{\alpha_i\}_{i=1}^{N}$ are impartitionable, then for any $t > 0$ the numbers in $\{t \alpha_i\}_{i=1}^{N}$ are also impartitionable. From~\eqref{eq:cosProd} we have:
				\be
					\ba
						\int_0^T & \left[ \prod_{i = 1}^{N} \cos (\alpha_i t) \right] dt \\
						=& \int_0^T \left[ \frac{1}{2^N} \sum_{\upsilon \in \{\pm 1\}^{N}} \cos \left( \sum_{i=1}^{N} \upsilon_i \alpha_i t \right) \right] dt \\
						=& \frac{1}{2^N} \sum_{\upsilon \in \{\pm 1\}^{N}} \int_0^T \cos \left( \sum_{i=1}^{N} \upsilon_i \alpha_i t \right) dt \equiv \varpi(T).
					\ea
				\ee
				Let $\iota \equiv \bm{\delta_\Sigma} (\{\alpha_i\}_{i=1}^{N})$. Now, using~\eqref{eq:intTbound} we have:
				\be
					\ba
						\varpi(T) \leq \frac{1}{2^N} \sum_{\upsilon \in \{\pm 1\}^{N}} \frac{2}{\sum_{i=1}^{N} \upsilon_i \alpha_i} \leq \frac{2}{\iota},
					\ea
				\ee
				and similarly $-\frac{2}{\iota} \leq \varpi(T)$. Since $\frac{2}{\iota}$ is constant, we have $\lim_{T \to \infty} \frac{\varpi(T)}{T} = 0$, and thus~\eqref{eq:cosProdMA} holds.
		\end{enumerate}
	\end{proof}
\end{lemma}

From Lemma~\ref{lem:cosProds} we immediately get:
\begin{cor}
	\label{cor:cosProdAvg}
	If the real number in $\{g_i\}_{i=1}^{N}$ are impartitionable, then for any $\{c_i\}_{i=1}^{N}$:
	\begin{enumerate}
		\item
			\be
				\label{eq:lemGamma2}
				\lim_{T \to \infty} \frac{1}{T} \int_0^T \prod_{i = 1}^{N} \left[ \cos^2(g_i t) + c_i^2 \sin^2 (g_i t) \right] dt = \prod_{i = 1}^{N} \frac{1 + c_i^2}{2},
			\ee
		\item
			\be
				\label{eq:lemB2}
				\lim_{T \to \infty} \frac{1}{T} \int_0^T \prod_{i = 1}^{N} \left[ 1 - c_i^2 \sin^2(g_i t) \right] dt = \prod_{i = 1}^{N} \left(1 - \frac{c_i^2}{2} \right).
			\ee
	\end{enumerate}
	
	\begin{proof}
		Let us first note that if the numbers in $\{g_i\}_{i=1}^{N}$ are impartitionable, then also the numbers in $\{2 g_i\}_{i=1}^{N}$ are impartitionable.
		
		We have $\cos^2 \alpha + c^2 \sin^2 \alpha = \frac{1}{2} (1 + c^2) + (c^2 - 1) \cos (2 \alpha)$. Thus the product within the integral in~\eqref{eq:lemGamma2} is equal to
		\be
			\label{eq:proofLemGamma2}
			\sum_{\sigma \subseteq \{1, \dots, N\}} \left[ \left( \prod_{i \in \sigma} \frac{1 + c_i^2}{2} \right) \left( \prod_{i \notin \sigma} (c_i^2 - 1) \cos (2 g_i t) \right) \right].
		\ee
		We can apply~\eqref{eq:cosProdMA} to see that the only term in~\eqref{eq:proofLemGamma2} which do not vanish in the limit is the one with $\sigma = \{1, \dots, N\}$, thus giving~\eqref{eq:lemGamma2}.
		
		Similarly, using $1 - c^2 \sin^2 \alpha = 1 - \frac{c^2}{2} - \frac{c^2}{2} \cos (2 \alpha)$ we get~\eqref{eq:lemB2}.
	\end{proof}
\end{cor}

In order to complete the proof of the Proposition~\ref{prop:timeAvg} using the Corollary~\ref{cor:cosProdAvg}, from the definition of $\gamma(t)$ we take $c_i^2 = \zeta_i^2$ and apply~\eqref{eq:lemGamma2} to get~\eqref{eq:gamma2TimeAvg}. Similarly to get~\eqref{eq:B2TimeAvg} we take $c_i^2 = (2 \lambda_i - 1)^2 \sin^2 \beta_i$ in~\eqref{eq:lemB2}.

\end{appendix}


\begin{thebibliography}{99}
\bibitem{ineq} Let $\gamma(t)=\mathrm{tr}(\varrho U_i^\dagger U_j)\equiv\mathrm{tr}(\sqrt\varrho V \sqrt\varrho)\equiv\mathrm{tr} X$, where we introduced $V\equiv U_i^\dagger U_j$ and $X\equiv\sqrt\varrho V \sqrt\varrho$.
	On the other hand $B(t)=\mathrm{tr}(\sqrt{U_i\sqrt\varrho U_i^\dagger U_j \varrho U_j^\dagger U_i\sqrt\varrho U_i^\dagger})=\mathrm{tr}(\sqrt{\sqrt\varrho V \sqrt\varrho\sqrt\varrho V^\dagger \sqrt\varrho})=\mathrm{tr}\sqrt{XX^\dagger}$.
	It is now an easy consequence of the polar decomposition that for any $X$, $|\mathrm{tr} X|\leq \mathrm{tr}\sqrt{XX^\dagger}$, from which $|\gamma(t)|\leq B(t)$.
\end{thebibliography}
\end{document}